\input{aipcheck}

\documentclass[
    final            
  ]{aipproc}

\layoutstyle{6x9}

\begin{document}

\title{Dust in Interstellar Clouds, Evolved Stars and Supernovae}





\author{T.~W.~Hartquist}{
  address={School of Physics and Astronomy, University of Leeds, 
	Leeds LS2 9JT, UK}
}

\author{S.~Van~Loo}{
  address={School of Physics and Astronomy, University of Leeds,
        Leeds LS2 9JT, UK}
}

\author{S.~A.~E.~G.~Falle}{
  address={Department of Applied Mathematical Sciences, University of Leeds,
	Leeds LS2 9JT, UK}
}

\author{P.~Caselli}{
  address={School of Physics and Astronomy, University of Leeds,
        Leeds LS2 9JT, UK}
}

\author{I.~Ashmore}{
  address={School of Physics and Astronomy, University of Leeds,
        Leeds LS2 9JT, UK}
}

\begin{abstract}
Outflows of pre-main-sequence stars drive shocks into molecular material
within 0.01 - 1 pc of the young stars. The shock-heated gas emits infrared
lines of $H_2$ and $H_2O$ and millimeter and submillimeter lines of many
species including $CO$, $SiO$, $H_2S$ and $HCO^+$. Dust grains are
important charge carriers and play a large role in coupling the magnetic
field and flow of neutral gas. Some understanding of the effects of the
dust on the dynamics of oblique shocks began to emerge in the 1990s.
However, detailed models of these shocks are required for the calculation of
the grain sputtering contribution to gas phase abundances of species
producing observed emissions. We are developing such models.

Some of the molecular species introduced into the gas phase by sputtering
in shocks or by thermally driven desorption in radiatively heated hot cores
form on grain surfaces. Recently laboratory studies have begun to contribute
to the understanding of surface reactions and thermally driven desorption
important for the chemistry of star forming clouds.

Dusty plasmas are prevalent in many evolved stars just
as well as in star forming regions. Radiation pressure
on dust plays a significant role in mass loss from some post-main-sequence
stars. The mechanisms leading to the formation of carbonaceous dust in
the stellar outflows are similar to those important for soot formation in
flames. However, nucleation in oxygen-rich outflows is less well understood
and remains a challenging research area.

Dust is observed in supernova ejecta that have not passed through the
reverse shocks that develop in the interaction of ejecta with ambient
media. Dust is detected in high redshift galaxies that are sufficiently
young that the only stars that could have produced the dust were so massive
that they became supernovae. Consequently, the issue of the survival of dust
in strong supernova shocks is of considerable interest. 
\end{abstract}



\maketitle

\section{INTRODUCTION}

   The claim that ``The presence of dust particles implies that the physical
state of such a medium may be somewhat different than that discussed in the
classical work by Eddington, in which only atoms were assumed to be
abundant'' appears prominently on the first page of Spitzer's important May
1941 paper [1]. In it Spitzer showed how to calculate the average charge on
an interstellar dust grain. The field of dusty astrophysical plasmas has
existed for close to seven decades. Here we describe some studies in a few
areas of dusty astrophysical plasmas beyond the Solar System.

\section{SHOCKS IN MAGNETIZED STAR FORMING REGIONS}

Stars form in regions in which the number density of hydrogen molecules is
at least $10^4$ cm$^{-3}$. Grains with a range of sizes between about 10 and
100 nm contain roughly one percent of the mass. The fractional ionizations
are $10^{-7}$ and lower. Grains are negatively charged and
their collisions with neutrals are important in coupling neutral
species to the magnetic field. Two-fluid [2] and multifluid models
of shocks driven into star forming regions by outflows of the young stars
have been constructed for more than 25 years.

   In such models of perpendicular shocks with speeds of less than about
40 km s$^{-1}$, the flows in all fluids are continuous and charged species are
accelerated earlier than neutrals, leading to dissipation regions that are
large compared to the mean free path. Many researchers have used a
rather crude treatment of the dynamics of charged species in such shocks.
In fact, dust grains in each of a number of size ranges and charged
states, electrons and gas phase ions should be treated as distinct fluids
[3,4].

   Wardle [5] showed that in a steady multifluid hydromagnetic model of a
fast-mode oblique shock, integration in the downstream direction does not
reach the downstream state because it corresponds to a saddle point. In
some cases, local equilibrium may not obtain everywhere, and integration in the
upstream direction is not always appropriate. Falle [6] performed the first
time-dependent study of plane-parallel, oblique shocks in which the dynamics
of all grain and gas phase species are treated rigorously, and all
three components of the current are calculated. Like Wardle [5], Falle [6]
made simple assumptions about the fractional ionization and charges  
and temperatures of the various species. We are
including a more thorough treatment of these quantities similar to that of
Pilipp et al. [3]. This is necessary for the calculation of dust sputtering
rates [7] important for the interpretation of molecular line data.

   Future models of evolving dense cores should include at
least two grain fluids even when strong shocks are not present.  

\section{SURFACE PROCESSES IN STAR FORMING REGIONS}

    Most gas phase species more massive than helium deplete onto dust grains
in 10 K objects, called dense cores, that are the precursors of stars. For a
molecular hydrogen number density of $10^5$ cm$^{-3}$, the depletion time is
roughly $10^5$ years which is comparable to the free fall time. Many species
accreted onto the dust grains react with one another creating hydrogenated
species like water and methane and some more complicated molecules
including methanol. The chemistry on the surfaces and the mechanisms
that desorb species are important because they affect the gas phase
abundances of molecules observed to probe the dynamics of star formation.
Desorption can be driven thermally which is particularly important in
hot cores, which have temperatures of 10$^2$ K, in regions where the birth 
of massive stars occurs. It can also be
induced by the absorption of photons [8,9], some of which cosmic rays
produce [10], and by exothermic reactions [11].

    A lack of detailed knowledge of binding energies and surface sweeping
rates of many species and of many reaction barriers hinders
the study of the grain surface chemical kinetics. It is made
even more difficult by the fact that usually only a small number of reactive
molecules are on a grain's surface, causing standard rate
equation treatments to be inappropriate. Green et al. [12] adopted a master
equation approach to study surface chemistry, but it is too
expensive computationally for most cases. Barzel and Biham
[13] have introduced a treatment based on the solution of moment
equations derived from the master equation. It is computational tractable
for systems of interest.

   Williams et al. [14] have reviewed theoretical and laboratory work on
$H_2$ formation on surfaces. Recent
studies have provided the first insight into the internal excitation
of $H_2$ upon formation on and escape from surfaces. Some laboratory
results on excitation were achieved through the use of state selective laser
induced ionization. Theorists have investigated the internal excitation of
$H_2$ when it is formed by the Eley-Rideal and Langmuir-Hinshelwood
mechanisms. The first involves the reaction of an atom just approaching
the encountered surface and reacting with a chemisorbed atom. The second
involves the reaction between two physisorbed atoms which sweep the surface.

   Williams et al. [14] have also reviewed recent work on laboratory studies
of adsorption and desorption on surfaces. The combination of reflection
absorption infrared spectroscopy (RAIRS) and temperature programmed
desorption (TPD) have enabled the development of a multi-step picture of
the carbon monoxide from a mixture of carbon monoxide and water ices
on surfaces. If the initial ice thickness is less than 1 nm, as the
temperature increases, CO desorbs at four distinct temperatures. RAIRS is
being used to study the yields of reactions on mimics of interstellar
grain surfaces.    

\section{DUST FORMATION IN AGB STARS}

Asymptotic Giant Branch stars burn helium and hydrogen in shells surrounding
carbon-oxygen cores. An AGB star's luminosity can approach 10$^4$ that of the
Sun and peaks in the red. Its atmosphere extends to about an AU. Mass loss
from an AGB star is driven by stellar pulsations and
radiation pressure on dust grains, and mass loss rates reach up to 10$^{-4}$
solar masses per year. The outflow speeds are 20 to
40 km s$^{-1}$. M-type AGB stars have the most oxygen-rich atmospheres.
S-type AGB stars are more evolved, and their atmospheres contain elements
produced by s-process nuclear burning and dredged up from lower layers.
C-type AGB stars are even more evolved and have carbon-rich atmospheres.
Ferrarotti and Gail [15] have modelled dust production in AGB stars as they
evolve from M-type to C-type. Dust forms in regions in which the number
density of hydrogen nuclei is 10$^9$ cm$^{-3}$.

Cherchneff [16] summarized the mechanisms leading to the formation of
carbonaneous dust grains in
carbon-rich stars. Reactions of simple species with acetylene and with atomic
hydrogen initiate the formation of ring molecules. Subsequent reactions
result in the generation of PAHs. Coagulation of PAHs occurs, and platelets
emerge and then join together to produce soot-like particles.

Dust generation in oxygen-rich
environments is a more challenging problem and was reviewed in 2004 by
Patzer [17]. He noted that ``In contrast to e. g. polyaromatic hydrocarbons
(PAHs) {\it no general building principles or schemes are known for small
oxide clusters}.'' Thus, searches for the minimum energy configurations,
based on the solution of Schroedinger's equation, for various combinations of
atoms have been made in order to identify stable cluster configurations.
However, full quantum chemical treatments cannot be used for large
systems. Consequently, semiempirical interaction potentials have sometimes been
employed. In recent studies of MgO clusters, Bhatt and Ford [18] have used a
combination of the compressible ion model and the polarizable ion model to
calculate interionic potentials.

The level of uncertainty that remains in this area is sufficient that Nuth
and Ferguson [19] felt compelled to entitle a paper ``Silicates Do Nucleate
in Oxygen-Rich Outflows: New Vapor Pressure Data for SiO''.

\section{DUST IN SUPERNOVAE}

Sugarman et al. [20] argued that up to 0.02 solar masses of dust formed in
Type II supernova 2003gd. Nozawa et al. [21] showed that the survival of
such dust as the ejecta pass through the reverse shock depends on the
supernova's energy, the dust particle size and the properties of the
envelope retained by the precursor star. They found that for some reasonable
sets of parameters the fraction, by mass, of the dust produced in the ejecta
that survives the reverse shock is as high as 0.8. 

{}

\end{document}